\documentclass[aps,prl,reprint,groupedaddress,showpacs,floatfix]{revtex4-1}

\newcommand{\EO}{E\"otv\"os} 
\usepackage{graphicx}
\usepackage[english]{babel}

\begin{document}


\title{Explanation of the EPF experiment in terms of gravity gradients}


\author{Gyula~T\'oth}
\affiliation{Department of Geodesy and Surveying,\\
Budapest University of Technology and Economics,\\
M\H uegyetem rkp. 3, Budapest, 1111, Hungary 
}


\date{\today}

\begin{abstract}
Results of the \EO\, Pek\'ar and Fekete (EPF) equivalence test were used by Fischbach and coworkers in 1986 as an argument in favor of a hypothetical fifth force. Although this hypothesis was abandoned in view of the negative experimental results that followed, we still miss plausible explanation of the EPF results. This situation motivated us to investigate the EPF test in terms of gravity gradients. 
This paper presents arguments that the results can be explained as a classical systematic effect related to the ambient gravity field.  We found that gradients of the ambient gravity field caused a false equivalence violation signal. 
Firstly, this was due to the time variation of gravity gradients, in spite of the fact that the experimenters were aware of it and designed a method to cancel it. Second, the EPF samples had different shapes and therefore the gravitational force was necessarily different in a constant, but inhomogeneous gravity field. We demonstrate that there is an ambient gravity field where these effects can fully reproduce the EPF results.
\end{abstract}



\maketitle


The \EO\, Pek\'ar and Fekete (EPF) equivalence test of gravitational and inertial masses of a body was an outstanding achievement in physics at the beginning of the 20th century \cite{eotvos_beitrage_1922}. They improved on the precision of previous tests by more than three orders of magnitude and were the first to use a torsion balance to test the equivalence principle. They found no violation on the level of accuracy 1/100,000,000 \cite{bod_one_1991}.

All of this was only of historical interest until 1986, when Fischbach and his coworkers reanalyzed the EPF data and found a composition dependence in terms of baryon number-to-mass ratios of the samples \cite{fischbach_reanalysis_1986,fischbach_long-range_1988}. They hypothesized a composition dependent fifth force.  A series of novel gravitational experiments followed, most notably from the E\"ot-Wash group \cite{stubbs_search_1987,stubbs_limits_1989,adelberger_testing_1990,nelson_1990,su_new_1994,gundlach_short-range_1997,smith_short-range_1999,wagner_torsion-balance_2012}, and found no evidence of such a fifth force. The original hypothesis in lack of experimental support was abandoned. There remained valid questions, however, about the EPF experiment.  The EPF correlation, in spite of every effort, has not yet been explained in terms of conventional or unconventional physics \cite{fischbach_long-range_1988,franklin_rise_2016}.

This situation motivated us to investigate the role of gravity gradients in the EPF test. Also we were motivated by our experience with the torsion balance in a nonlinear gravity field \cite{csapo_2009} and we asked if any such effect might be visible in the results of the EPF experiment.

\begin{figure}[h!]
\centering
\includegraphics[scale=0.31]{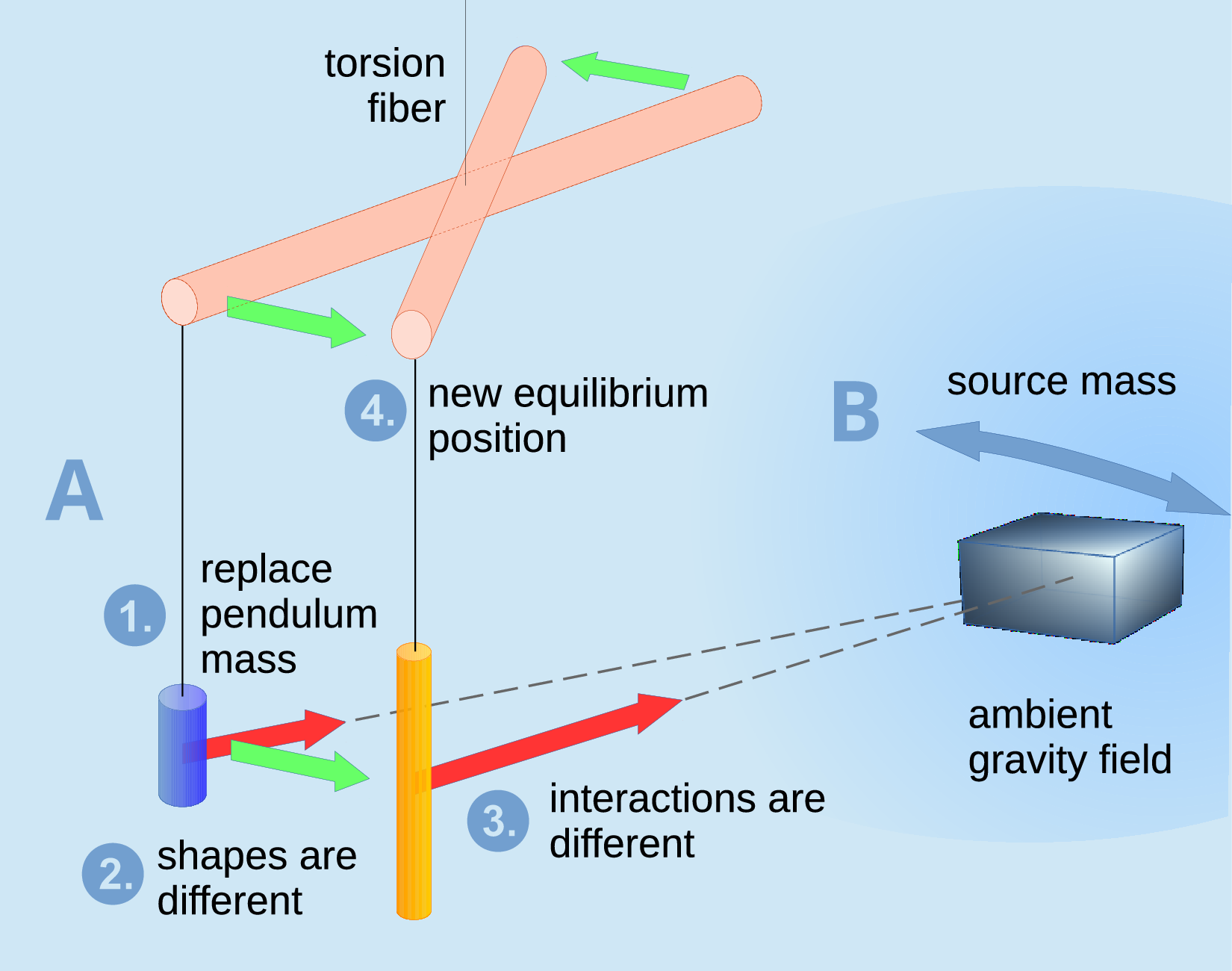}
\caption{A. In the EPF experiment lower mass of the balance was replaced with different samples. Sample geometry variation changed the coupling with the ambient gravity field and lead to variation of the direction of gravitational force. This caused the balance to move into a new equlibrium position, even when the equivalence principle was not violated and the ambient gravity field was unchanged. B. Gravity field might change during the experiment }
\label{fig:balance}
\end{figure}

In this paper we present arguments that the EPF results can be explained as a classical systematical effect related to the ambient gravity field. We found by using source mass modeling of the ambient gravity field that gravity gradient effects themselves are enough to fully reproduce the EPF results without any fifth force effect. The essential aspects of our analysis are summarized on Fig.~\ref{fig:balance}.

In the following first we emphasize briefly relevant details of the EPF measurements. Then we introduce the formalism which describes the interaction between the balance and the ambient gravity field. Next, a measurement model for the \EO\ parameter is derived that describes a possible violation of the equivalence principle. Then we explain construction of the ambient gravity field model and present results computed by our model. Finally, we discuss the results and point out some important conclusions.

The purpose of the EPF experiment was to compare gravitational acceleration due to the Earth on different materials or samples \cite{eotvos_beitrage_1922}. There were 10 pairs of such samples. The effect on the samples below the arm was compared to the fixed upper mass by the \EO\ parameter $\eta$. This parameter is the ratio of the horizontal component of the differential acceleration of the upper and lower masses and the horizontal component of the gravitational acceleration \cite{Adelberger:2009zz}. The results of EPF tests were finally described in terms of variation of the \EO\ parameter $\Delta\eta$ between different pairs of samples.

When we analyze the EPF experiment from the point of view of possible gravity gradient effects, it is important to realize that they used two essentially different methods: Method 2 and Method 3 for 5 pairs of samples each. For what they called “Method 2”, measurements with the compared samples were not simultaneous. Due to the time delay between measurements of samples of the pair, ambient gravity field changes would give a false violation signal. To get rid of this effect, in case of their more advanced Method 3, they measured different sample pairs at the same time with a double balance to avoid time delay. Measurement of the sample pair was repeated after exchanging samples between balances of the double balance. This procedure was necessary to correct for the small effect of azimuth differences of the balance arms.

To formulate the effect of gravity gradients on $\Delta\eta$ we used multipoles \cite{durso_translation_1997}. Multipoles proved to be useful to describe gravitational interaction between the masses of the torsion balance and the masses outside that produce the ambient gravity field \cite{su_new_1994}. The ambient field is characterized by $Q_{lm}$ multipole fields; with this characterization the gravitational torque on the balance is
\begin{equation}
T_g = -\frac{\partial W}{\partial \phi} = -4\pi iG \sum_{l=2}^\infty \frac{1}{2l+1} \sum_{m=-l}^l m \; q_{lm} Q_{lm} e^{-im\phi}  . 
\label{eq:Tg}
\end{equation}

Here $W$ is gravitational potential energy, $G$ is universal constant of gravitation, $q_{lm}$ are multipole moments of the balance calculated in a body-fixed frame,  $\phi$ is azimuth of the balance's arm and star denotes complex conjugation. Azimuth is measured from the $x$-axis, positive towards the $y$-axis. The axes of our system are as follows: the $x$ axis points to North, $y$ to East and $z$ to Down. No torque is produced by the $Q_{11}$ multipole field, because the arm is hanging freely on the torsion fiber; hence the sum starts from $l = 2$.

When the \EO\ parameter variation is not zero, the equivalence principle is violated. The \EO\ parameter variation is defined for Method 2 by Eq.~(\ref{eq:Method2}), 
\begin{equation}
\Delta\eta=c\left ( \frac{v}{w} - \frac{v'}{w'} \right ) 
\label{eq:Method2}
\end{equation}
where  $v$ and $w$ denote deflection differences of the balance arm in E-W and N-S directions, and by Eq.~(\ref{eq:Method3})
\begin{equation}
\Delta\eta=\frac{c}{2} \left [   \left ( \frac{v_1}{w_1} - \frac{v_2'}{w_2'} \right ) + \left ( \frac{v_2}{w_2} - \frac{v_1'}{w_1'} \right ) \right ]
\label{eq:Method3}
\end{equation}
for Method 3, where primes indicate a different sample and subscripts denote individual balances of the double balance. The proportionality constant is
\begin{equation}
c= \frac{w\tau}{4LM_a l_a C\sin\varphi},
\end{equation}
where $L$ is distance to the scale in scale units, $M_a$ is mass of the sample, $l_a$ is length of the balance arm, $\tau$ is torsion constant of the fiber and $C \sin\varphi$ is the horizontal component of centrifugal acceleration. Small variations of the \EO\ parameter can also be caused by azimuth differences of the arm; we omitted these since we were interested in gravity gradient effects.

In the second step, we expressed the ratio of $v$ and $w$ in terms of gravitational torque differences, which in turn are related to field multipoles according to Eq.~(\ref{eq:Tg}). We assumed a symmetrical mass distribution of the balance with respect to the plane of the arm's axis and the fiber. In this case all $q_{lm}$ are real and
\begin{equation}
\frac{v}{w} = -\frac{\text{Re}(p)}{\text{Im}(p)},
\label{eq:vwequ}
\end{equation}
where 
\begin{equation}
p = q_{21}Q_{21} + \frac{5}{7}\; q_{31}Q_{31} - \frac{5}{7}\; q_{33}Q^*_{33} + \frac{5}{9}\; q_{41}Q_{41} - \frac{5}{9}\; q_{43}Q^*_{43} .
\label{eq:pequ}
\end{equation}

The $–v/w$ ratio is tangent of the complementary phase angle of the complex number $p$. The phase of $p$  is thus linked to the $v/w$ ratio, which is in turn proportional to the \EO\ parameter variation $\Delta\eta$. Equation (\ref{eq:pequ}) shows that $p$ is a linear combination of multipole moment and multipole field products $q_{lm}Q_{lm}$  with odd order. To summarize, Eqs. (\ref{eq:Method2}--\ref{eq:pequ}) give formulation of the effect of gravity gradients on the output of the EPF experiment up to $l \le 4$. 

This formulation reveals important things from the point of view of ambient gravity gradients. First, if we neglect all higher degree terms of $p$, any false violation signal must come from the change of $Q_{21}$. Change of $q_{21}$  cannot change the phase of $p$, because $q_{21}$ is real. The possibility of false violation by a changing $Q_{21}$ was considered by EPF, and they designed Method~3 to avoid that. 

But there is a second possibility. When higher-degree terms are not negligible and phases of field multipoles are different, then changes in $Q_{lm}$ and/or $q_{lm}$  modify the phase of $p$ and lead to a false violation signal. We emphasize this point: even if \emph{all} $Q_{lm}$'s are constant, that is we have a time invariable ambient gravity field, variation of $q_{lm}$ multipole moments generally gives rise to a false violation signal. Therefore this signal must appear even in the results of Method~3. 

Finally, in case of $q_{lm}$ variation, the false violation signal depends also on $Q_{lm}$ as their \emph{product} is what counts. So this signal must have a gravity field dependence. Particularly, this affects results of Method 3. If ambient gravity field varies between two such measurements with different pairs of samples, results of Method 3 will also vary. 

The samples used in the EPF experiment were all cylinders suspended vertically but their shapes were different. Shape dependence is critical, since it affects the following low degree multipole moments ($l = 2,3,4$):
\begin{eqnarray*}
&& q_{20} = \frac{1}{24} \sqrt{\frac{5}{\pi}} M (H^2 -3R^2-6x^2-6y^2+12z^2)  \\
&& q_{31} = \frac{1}{8} \sqrt{\frac{21}{\pi}} M (x-iy)(H^2 -3R^2-3x^2-3y^2+12z^2)  \\
&& q_{41} = -\frac{3}{8} \sqrt{\frac{5}{\pi}} M z (x-iy)(H^2 -3R^2-3x^2-3y^2+4z^2)
\end{eqnarray*}
where $M$ is mass of the cylinder, $R$ is radius, $H$ is height, $x$, $y$, $z$ are coordinates of the center of mass of the cylinder. Shape dependencies of $q_{31}$ and $q_{41}$ according to Eq.~(\ref{eq:pequ}) must appear in the EPF results.

\begin{figure}[h!]
\centering
\includegraphics[scale=0.65]{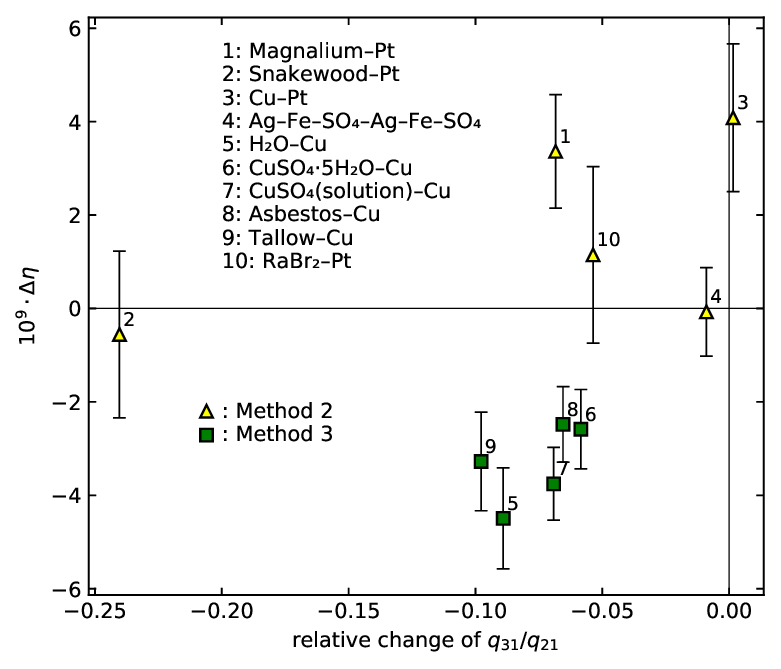}
\caption{Variation of the \EO\  parameter, $\Delta\eta$ is shown here as a function of relative change of the ratio $q_{31}/q_{21}$ of multipole moments of the balance between two samples. Linear dependence is expected in a steady-state ambient gravity field. The figure shows that this is approximately true only for Method~3 results, which are less sensitive to time variation of the gravity field. This indicates time variation of the ambient gravity field during the experiment.}
\label{fig:dq23}
\end{figure}

To summarize, gravity gradients of the ambient gravity field might have caused a false equivalence violation signal in the EPF experiment. First, it was because of time variation of gravity gradients. Even in a steady-state ambient gravity field there was a false violation signal. It was from the change of coupling to the gravity field as a function of sample geometry (see Fig.~\ref{fig:balance}). EPF used samples of different shape, hence this effect was necessarily non-null in the experiment. Even with their Method 3 this effect was non-null and is sensitive to the ambient gravity field variation.

We needed both multipole moments and field multipoles to calculate variation of the \EO\ parameter $\Delta\eta$. It was straightforward to calculate multipole moments of the balances used by EPF from the parameters published in the EPF paper \cite{eotvos_beitrage_1922}, by using closed expressions of inner multipole moments \cite{durso_translation_1997,Stirling:2017rgy}. Unfortunately, multipole fields have not been measured during the experiment and are basically unknown. On the contrary, multipole fields were carefully measured and compensated in rotating torsion balance tests \cite{su_new_1994,wagner_torsion-balance_2012,xu_2017} to minimize any possible false gravity gradient violation signal.

\begin{figure}[h!]
\centering
\includegraphics[scale=0.48]{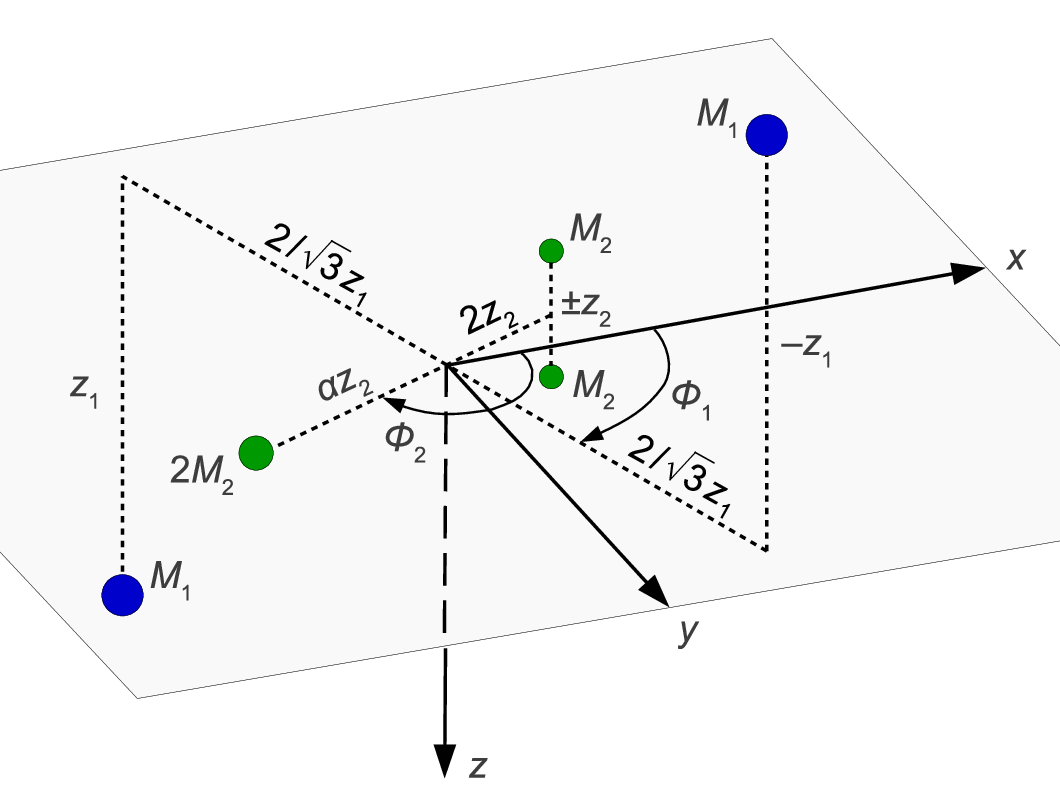}
\caption{Point mass model of the ambient gravity field. The first part contains two point masses $M_1$ and $M_1$ symmetrically placed around the origin. It models the $Q_{21}$ field multipole and has a non-zero $Q_{43}$. Three point masses $2M_2$, $M_2$ and $M_2$ form the second part to model the $Q_{31}$ field multipole (see Table \ref{tab:pmmtable}). Parameter $\alpha$ is $5/2 \sqrt[8]{4/5}$.}
\label{fig:ambient}
\end{figure}

To circumvent the difficult problem of missing field multipoles we constructed a simple source mass model of the ambient gravity field (Fig.~\ref{fig:ambient}). This simple model has shown that a particular ambient gravitational field can reproduce the measured effects. The largest effect is expected from the two lowest degree $Q_{21}$ and $Q_{31}$ multipole fields according to Eq.(\ref{eq:pequ}). Only $q_{31}$ and $q_{41}$ depend on the shape in case of a vertical cylinder. The largest shape effect is expected from $Q_{31}$ because Eq.~(\ref{eq:Tg}) converges as $(r/R)^l$ where $r$ is a typical dimension of the torsion balance and $R$ is a characteristic distance from the pendulum to the closest source \cite{su_new_1994}.

\begin{table*} 
\caption{Details of the ambient gravity field model of low order ($l \le 4$) $Q_{lm}$ field multipoles. The table contains horizontal distances, azimuths, $z$-coordinates of the point masses as well as calculated non-zero field multipoles of the model. \label{tab:pmmtable}}
\begin{ruledtabular}
\begin{tabular}{cccccc}
part & mass & horizontal distance & azimuth & $z$-coordinate & non-zero field multipoles \\
\hline
  1. & $M_1$ & $2/\sqrt{3}z_1$ & $\Phi_1$      & $z_1$  & $Q_{21}=-9/343 \sqrt{210/\pi} M_1 e^{i\Phi_1}/z_1^3$ \\
     & $M_1$ & $2/\sqrt{3}z_1$ & $\Phi_1 +\pi$ & $-z_1$ & $Q_{43}=-162/240 \sqrt{5/\pi} M_1 e^{i3\Phi_1}/z_1^5$ \\
\hline
  2. & $2M_2$ & $5/2\sqrt[8]{4/5}z_2$ & $\Phi_2$      & $0$    &  \\
     & $M_2$  & $2z_2$         & $\Phi_2+\pi$  & $z_2$  & $Q_{31}=2/625 \sqrt{105/\pi} M_2 e^{i\Phi_2}/z_2^4$ \\
     & $M_2$  & $2z_2$         & $\Phi_2+\pi$  & $-z_2$ & 
\end{tabular}
\end{ruledtabular}
\end{table*}

Our mass model consisted of 5 point masses. The number of independent parameters is 4. Details of the model are found in Table \ref{tab:pmmtable}. We mention that the $v/w$ ratio computed from the model must conform to the EPF measurements. Hence, the azimuths $\Phi_1$ were constrained to yield the measured $v/w$ ratios.

\begin{figure}[]
\centering
\includegraphics[scale=0.65]{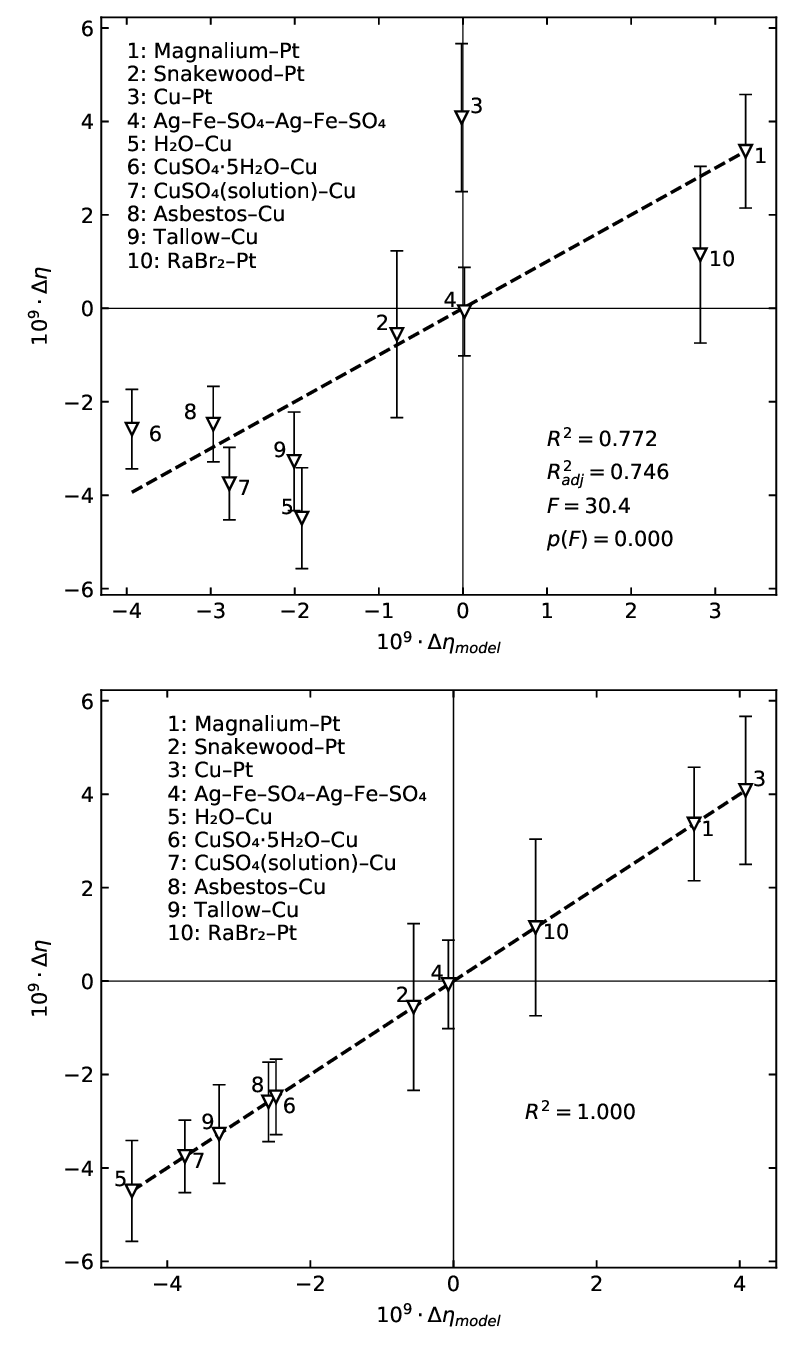}
\caption{These two figures show two extreme cases of correlation of our ambient gravity field bias model with the EPF experiment in terms of modeled and measured $\Delta\eta$ \EO\  parameter differences. The upper figure shows results of Case~1. In this case no variation of the ambient gravity field was allowed. Consequently,  modeled $\Delta\eta_{model}$ was due to varying sample geometry alone. The lower figure shows perfect correlation in Case~2. This fit was achieved by allowing small variations of the ambient gravity field model between measurements that were not taken at the same time (See Fig. \ref{fig:EPFparam}). Although it is unreasonable to require a perfect fit, it demonstrates clearly that the original EPF measurements can be interpreted fully as a false gravity gradient effect.}
\label{fig:pmmfits}
\end{figure}

The parameter space of the model was searched for optimum solutions by differential evolution~\cite{Storn1997}. Optimality criterion was that the sum of weighted squared differences between model $\Delta\eta_{model}$ and measured  $\Delta\eta$ should be minimum. Weights were assigned from the standard deviations of the results. More precise values of \EO\ parameter variations $\Delta\eta$ for the 10 sample pairs were recalculated from the original data \cite{eotvos_beitrage_1922}.

First we checked sample geometry dependence of the EPF results. This was done by computing relative changes of the ratios of multipole moments,  $q_{31}/q_{21}$ and $q_{41}/q_{21}$. It can be demonstrated by Eqs.(\ref{eq:Method2},\ref{eq:Method3},\ref{eq:vwequ},\ref{eq:pequ}) that this ratio in a steady-state ambient gravity field is approximately in linear correlation with the \EO\  parameter variation $\Delta\eta$. Figure~\ref{fig:dq23} shows the correlation with computed relative changes of the $q_{31}/q_{21}$ ratio between the samples of the same sample pair. 

\begin{figure}[t!]
\centering
\includegraphics[scale=0.59]{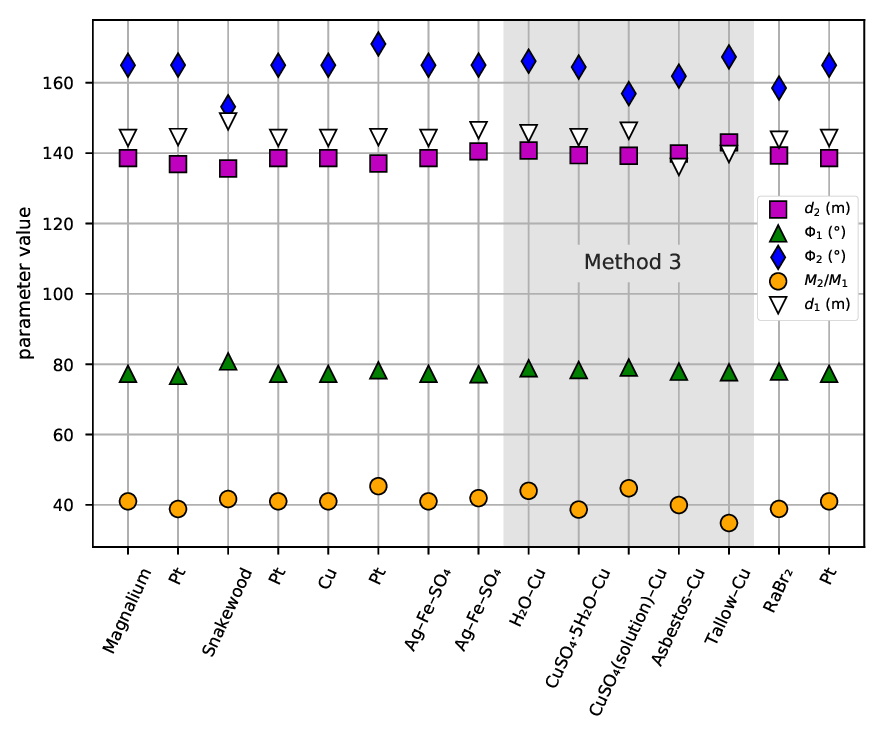}
\caption{The figure shows parameter values of the ambient gravity field model that were required for exactly reproducing the EPF results (See lower sub-figure of Fig. \ref{fig:pmmfits}). Shading indicates measurements with Method~3. For each sample pair measured with Method~2 there were two models since these measurements were not taken at the same time. The ambient gravity field model was a simple 5 point mass model composed of a 2-point $Q_{21}$,$Q_{43}$ and of another 3-point $Q_{31}$ field. Parameters $d_1$ and $\Phi_1$ resp. $d_2$ and $\Phi_2$ are horizontal distance and azimuth belonging the 2-point resp. 3-point mass models. $M_2/M_1$ is the mass ratio of the two models. 84\% of the relative parameter changes are below $\pm$5\%, the maximum is 17.9\%. These changes seems reasonable since EPF reported on the construction of a nearby building during the observations. }
\label{fig:EPFparam}
\end{figure}

Next, we modeled the ambient gravity field with the presented 5-point mass model. We assumed a relatively strong source of $Q_{31}$ field multipole at 20~m characteristic distance. It was because about 20~meters from the measurement site there was a strong concrete tower as reported by \cite{bod_one_1991}. We considered two extreme cases: in Case~1, no variation of the model was allowed. Upper sub-figure of Fig.~\ref{fig:pmmfits} shows the correlation between this model and the original EPF measurement in terms of variations of the \EO\  parameter  $\Delta\eta$. In Case 2, small variations were allowed for parameters of the mass model. Lower sub-figure of Fig. \ref{fig:pmmfits} shows that this way a perfect match between the model and the original EPF measurement could be achieved in terms of variations of the \EO\  parameter  $\Delta\eta$.

Figure \ref{fig:EPFparam}. presents model parameters required for the perfect fit in Case 2. This fit was achieved with only a 2.6\% average absolute variation of the mass model's parameters. Additionally, 84\% of the relative parameter changes were below $\pm$5\%. Maximum parameter variation was 17.9\%, and the three largest variations were found in the mass ratio parameter $M_2/M_1$.

Figure~\ref{fig:dq23} shows no linear dependence for results by Method~2; on the contrary, results by Method~3  show an approximate linear dependence. This is not surprising, since time variation of the ambient gravity field may easily hide the effect of sample geometry. On the other hand, Method~3 results are less sensitive to time variation of the gravity field, and thus the sample geometry effect becomes visible. This indicates that the ambient gravity field and its variation during the experiment must have played a significant role. In case of purely random effects no such distinction should be seen between the two methods.

Interpretation of the results obtained by source mass modeling confirm the role of time variation of the ambient gravity field during the EPF experiment. When no time variation was allowed, we found moderate correlation between modeled and measured \EO\  parameter differences $\Delta\eta$. Even for this fit quite unrealistic model parameters (too large $M_2/M_1$ ratio and too small $d_2$) were required.

On the other hand, when time variation of the source mass model was allowed, we got reasonable results. Although the assumption of a perfect fit without any statistical fluctuation is unrealistic, Fig.~\ref{fig:EPFparam} shows that both magnitude of calculated parameters of the mass model and range of their variations are feasible. These variations must be taken into account to explain the EPF results. In connection with these we mention, that EPF reported on the construction of a nearby building during the observations \cite{selenyi_roland_1953}. This construction work may explain variation of the ambient gravity field.

In conclusion, we are confident that our findings provide a possible explanation of the EPF data as a systematic effect coming from the ambient gravity field. The EPF results were not truly random as the experimenters expected, but were infected by a systematic error. In reality the effect is not material (composition) dependent; the correlation found by Fischbach and his coworkers with baryon number-to-mass ratios of the samples may be viewed as accidental in light of the present results. We propose experimental verification of the gravity gradient effect by using an original torsion balance.

\begin{acknowledgments}
The author thanks colleagues at Wigner Research Centre for Physics for discussions, especially P\'eter V\'an for calling our attention to the EPF experiment, Lajos V\"olgyesi and Gy\"orgy Szondy for their encouragement and support. The author is grateful to J\'ozsef Cserti for pointing to an error in the calculation of multipoles of the point mass model.
\end{acknowledgments}

\end{document}